\documentclass[twocolumn,appendixfloats]{aastex6}

\usepackage{graphicx}
\usepackage{epstopdf}
\usepackage{natbib}
\usepackage{subfigure}
\usepackage{amsmath}

\begin{document}

\title{A subtle IR excess associated with a young White Dwarf in the Edinburgh-Cape Blue Object Survey}

\author{E. Dennihy \altaffilmark{1}, John H. Debes \altaffilmark{2}, B. H. Dunlap\altaffilmark{1}, P. Dufour \altaffilmark{3}, Johanna K. Teske\altaffilmark{4,}\altaffilmark{5}, J.C. Clemens \altaffilmark{1}}
\affil{\altaffilmark{1}Physics and Astronomy Department, University of North Carolina at Chapel Hill, Chapel Hill, NC 27599; edennihy@unc.edu, clemens@unc.edu, bhdunlap@unc.edu \\ \altaffilmark{2}Space Telescope Science Institute, Baltimore, MD 21218; debes@stsci.edu \\ \altaffilmark{3}Institiut de Recherche sur les Exoplan\'etes (iREx), D\'epartement de Physique, Universit\'e de Montr\'eal, Montreal, Qu\'ebec H3C 3J17, Canada; dufourpa@astro.umontreal.ca \\ \altaffilmark{4}Carnegie DTM, 5241 Broad Branch Road, NV, Washington, DC 20015, USA; jteske@carnegiescience.edu \\ \altaffilmark{5}Carnegie Origins Fellow, jointly appointed by Carnegie DTM \& Carnegie Observatories}

\begin{abstract}
We report the discovery of a subtle infrared excess associated with the young white dwarf EC\,05365--4749 at 3.35 and 4.6\,$\mu$m. Follow-up spectroscopic observations are consistent with a hydrogen atmosphere white dwarf of effective temperature 22\,800\,K and log [\emph{g} (\,cm\,s$^{-2}$) ] = 8.19. High resolution spectroscopy reveals atmospheric metal pollution with logarithmic abundances of [Mg/H] = --5.36 and [Ca/H] = --5.75, confirming the white dwarf is actively accreting from a metal-rich source with an intriguing abundance pattern. We find that the infrared excess is well modeled by a flat, opaque debris disk, though disk parameters are not well constrained by the small number of infrared excess points. We further demonstrate that relaxing the assumption of a circular dusty debris disk to include elliptical disks expands the widths of acceptable disks, adding an alternative interpretation to the subtle infrared excesses commonly observed around young white dwarfs.
\end{abstract}


\section{Introduction}

The discovery of infrared excesses around white dwarf stars and their interpretation as debris disks has opened an exciting new avenue for planetary research. Atmospheric metals in otherwise pristine hydrogen atmosphere white dwarf stars are in most cases now understood to be delivered to the surface from dust disks formed from tidally disrupted planetary bodies \citep{deb02, jur03, ver14}. The extreme stargazing orbits required for the tidal disruption of rocky exo-planetary material provide an interesting framework for studies on post-main sequence planetary evolution through planetary and binary interactions \citep{deb12, bon15}. Detailed abundance studies of white dwarf atmospheres have shown signatures of carbanaceous asteroids, planetary lithospheres, and even water \citep{jur15,far13,zuc11}. A subsample of dusty disk hosting white dwarfs with gaseous disk components have also shown dynamical evolution on timescales of years \citep{wil15}. These comprehensive studies offer insight into exoplanetary chemical composition and dynamics not accessible to other exoplanet studies. 

The systematic search for these ideal exoplanetary laboratories saw early progress with the targeted infrared photometry of known metal-polluted white dwarf stars \citep{kil06}. The \emph{Spitzer Space Telescope} was invaluable for confirming many disks and constraining their radii to within the tidal disruption radius of their host white dwarf stars \citep{jur07, far08}. Later, all sky near-infrared surveys such as the Two Micron All Sky Survey (2MASS; Skrutskie et al. 2006) and the \emph{Wide-field Infrared Sky Survey} (\emph{WISE}; Wright et al. 2010) provided a basis for rapidly surveying known white dwarfs for infrared excesses \citep{deb11, hoa13}. To date, these techniques have increased the sample of debris disk hosting white dwarfs to nearly 40 confirmed systems \citep{roc15}. The continued growth of the sample is critical for exploring the full range of phenomena that evolved planetary systems exhibit.   

In this paper we present the discovery of an infrared excess at the DA white dwarf EC\,05365--4749 (hereafter EC\,05365), as well as the detection of atmospheric metal pollution in its optical spectra. Originally identified in the first two zones of the Edinburgh-Cape Blue Object Survey \citep{kil97, odo13}, EC\,05365 is identified in several all sky photometric surveys at $5^{\rm h}37^{\rm m}53.5^{\rm s}$, -$47^{\circ}58'05.3''$ in FK5 coordinates. In section 2 we discuss the photometry and spectroscopic follow-up collected to construct an informative spectral energy distribution (SED) and model the atmospheric pollution. Section 3 begins by establishing the infrared excess, followed by a range of modeling efforts including relaxing the assumption of a circular debris disk to include elliptical debris disks. Section 3 closes with an analyses of the atmospheric pollution and accretion rates. Section 4 places EC\,05365 in the context of other dusty disk hosting white dwarfs and presents recommendations for follow up observations.  

\section{Target Selection and Observations}

Following the prescriptions laid out by \cite{deb11} for the \emph{WISE} Infrared Excess around Degenerates survey, we cross-correlated the DA atmosphere white dwarfs identified in Zones 1 and 2 of the Edinburgh-Cape Blue Object survey (hereafter EC survey) \citep{kil97, odo13} with the ALLWISE photometric catalog and 2MASS \citep{skr06} All Sky Data Release Point Source Catalog (2MASS PSC). Many but not all of the white dwarfs identified in the first two published zones of the EC survey have been incorporated into the McCook and Sion white dwarf catalog \citep{mcc99} and were included in the WIRED catalog by \cite{hoa13}. This search was carried out with the missing targets in mind. We will extend our search to include the remaining EC Survey zones as they are published. Of the approximately 110 targets surveyed in this paper, two were identified as strong excess candidates: EC\,05365, the focus of this paper, and WD 1150--153, a previously known dusty white dwarf \citep{kil07}. These numbers are roughly consistent with recent estimates of the frequency of dust disks around white dwarfs \citep{deb11, bar14, roc15}.

\subsection{Photometry}
We collected published photometric measurements of EC\,05365 from the 2MASS PSC \citep{cut03}, the ALLWISE photometric catalog \citep{cut13}, the AAVSO Photometric All Sky Survey (APASS) \citep{hen09}, the \emph{Galaxy Evolution Explorer} (\emph{GALEX}; Martin et al. 2005) Data Release 6, and the VISTA VHS 3rd data release \citep{mcm13}, enabling the construction of a spectral energy distribution spanning ultraviolet to near-infrared wavelengths. We also collected near-infrared images for astrometric information from the VISTA Science Archive, described in \cite{irw04}, \cite{ham08}, and \cite{cro12}. 

\begin{figure}
\begin{centering}
\includegraphics[scale=0.5,clip=True]{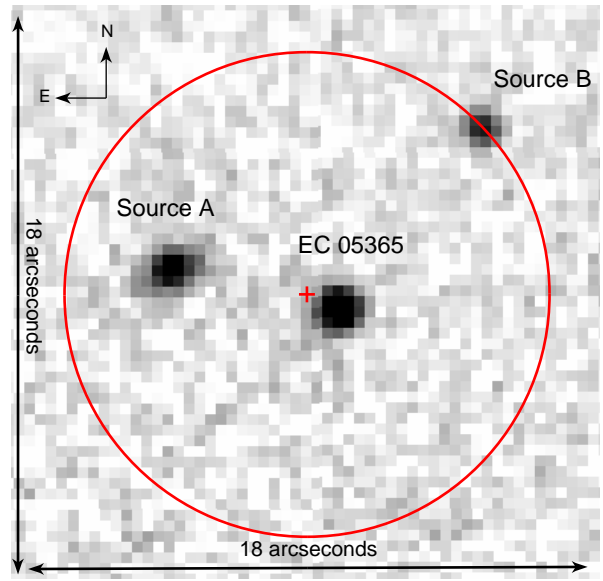}
\end{centering}
\caption{Infrared \emph{K$_{s}$} image taken from the VISTA VHS Survey DR3. EC\,05365 is identified at $5^{\rm h}37^{\rm m}53.5^{\rm s}$, -$47^{\circ}58'05.3''$ in FK5 coordinates along with a nearby sources `A' and `B'. A 7.8 arcsecond circle is shown centered on the red cross which identifies the ALLWISE detection photocenter.}
\end{figure}

\begin{deluxetable}{lccccr}
\tablecaption{VISTA VHS Photometry}
\tablehead{Object & \emph{J} (mag) & $\sigma_{\emph{J}}$ & \emph{K$_{s}$} (mag) & $\sigma_{\emph{K$_{s}$}}$ & Sep ($\arcsec$)} 
\startdata
EC\,05365 & 16.099 & 0.010 & 16.114 & 0.038 & 1.12 \\
Source A & 17.869 & 0.041 & 16.615 & 0.057 & 4.46 \\
Source B & 17.796 & 0.036 & 17.469 & 0.128 & 7.73 \\
\enddata
\end{deluxetable}

As discussed in \cite{deb11}, the large point-spread function of the \emph{WISE} telescope often leads to contamination or source confusion. High spatial resolution follow-up images are necessary for identifying nearby contaminants. The \emph{WISE} photometry system implements profile fit photometry and source deblending routines that are robust against contamination of resolved sources separated by more than 1.3$\times$FWHM (approx. 7.8 arcseconds for \emph{W1}) of the point-spread function of the band of interest, and is capable of flagging sources contaminated by a nearby unresolved source by virtue of the goodness of fit of the best fit point-spread profile. The use of these deblending routines in the photometric measurement of a source is captured in a set of photometric flags; `nb' for the passive deblending routine of resolved sources separated by more than 1.3$\times$FWHM of the point-spread function of the band, and `na' for the active deblending of a pair of unresolved sources. For the photometry taken at the source position associated with EC05365, the source was not flagged for active profile deblending, indicating that the reported photometry is not corrected for contamination for a nearby source within 7.8 arcseconds. 

Figure 1 displays the VISTA VHS \emph{K$_{s}$} band image which allowed us to identify two potential nearby contaminants, Source A and Source B. The red circle indicates the limit of source resolution implemented by the ALLWISE photometry pipeline prior to attempting blended profile fit photometry. The VISTA \emph{J} and \emph{K$_{s}$} band measurements and approximate separation from the ALLWISE detection photocenter for each source are given in Table 1. Source B is just at the limit of resolution. Given the relatively faint \emph{K$_{s}$} magnitude and \emph{J} - \emph{K$_{s}$} color, we suspect it is below the detection limit of the WISE bands and not a source of contamination for the ALLWISE detection. Source A however is comparable in \emph{K$_{s}$} brightness and has a \emph{J}-\emph{K$_{s}$} color indicating it could have a measurable flux in the \emph{WISE} 1 and 2 bands. Source A is also extended and has a high probability of being extra-galactic.Considering only the information contained in the ALLWISE catalog, we have no way of determining if the ALLWISE measurement represents a blend of the target and Source A, or if it is instead consistent with a single source at either the position of the target or Source A. However, given that we know the positions of the target and Source A very well (compared to the course sampling of the \emph{WISE} pixel scale) from the VISTA catalog, we can perform a set of estimations to determine the most likely case. We discuss this potential for contamination in detail in the context of our spectral energy distribution in Section 3.2.

\subsection {Spectroscopic Follow-up}
Photometry alone is sufficient to identify the infrared excess, but spectroscopic follow-up is necessary for identifying and characterizing the atmospheric pollutants that indirectly probe the remnant planetary system. We first observed EC\,05365 with the Goodman spectrograph \citep{cle04} on the SOAR telescope configured to search for atmospheric metal pollution on 2015 April 04. The 0.46'' long slit and 1800 l/mm grating were chosen to maximize resolution ($\approx$ 0.66\,\AA\, per resolution element) while allowing the spectral range to probe for the \ion{Ca}{2} H\,\&\,K transitions at 3934\,\AA\, and 3967\,\AA\,, and the \ion{Mg}{2} doublet at 4481\,\AA. We achieved a S/N per resolution element of 27 near 4000\,\AA\, in 3240s of combined exposure time. The data revealed an absorption feature at the approximate location of the \ion{Mg}{2} doublet with an equivalent width (EW) of 75 $\pm$ 16\,m\AA, though without a published radial velocity or suitable comparison line we could not rule out the possibility that the feature was interstellar.

To confirm the detection of Mg and search for additional metal species, we observed EC\,05365 with the MIKE spectrograph on the Magellan telescope on 2015 April 30. We used the 0.5''$\times$5'' slit which translated to a resolution of 0.083\,\AA\,per resolution element and a S/N per resolution element of 25 near 4000\,\AA\, in the combined 3$\times$1200\,s exposures. No standard star was observed as the goal was strictly to confirm atmospheric pollution measured relative to the local continuum. The data were extracted and flatfielded using the MIKE reduction pipeline, with methodology described in \cite{kel00} and \cite{kel03}. We confirmed the detection of \ion{Mg}{2} in the Goodman data with an EW of 83 $\pm$ 6\,m\AA\,  and discovered a Ca K absorption feature with an EW of 47 $\pm$ 11\,m\AA. Both features were found to be at the photospheric velocity as calibrated against the non-LTE core of the hydrogen alpha absorption line. After applying heliocentric corrections we found the photospheric velocity (combined gravitational redshift and radial velocity) to be 40 $\pm$ 7\,km\,s$^{-1}$.

Finally, to better constrain the atmospheric parameters we observed EC\,05365 with the Goodman spectrograph on 2015 May 21 with a low resolution setup. The spectrograph was configured with the 930 l/mm grating and the wide 3.0'' long slit to minimize the effect of variable seeing conditions which varied between 1.4-1.6''. The seeing limited resolution was $\approx$ 4.2\,\AA\, per resolution element with a wavelength range of 3700 to 5200 angstroms and we achieved a S/N of 110 per resolution element near 4200\,\AA\, in the combined 7$\times$180\,s exposures. We reduced and extracted the spectra with standard IRAF techniques and employed user developed techniques for wavelength and continuum calibration. The spectroscopic standard LTT 2415 was observed to provide continuum calibration but, as discussed in section 3.1, cloudy conditions rendered the standard unreliable for flux calibration. 

\section{Analysis and Discussion}

\subsection{Atmospheric Parameters}

\begin{figure}[t]
\includegraphics[angle=270,scale=1.0,clip=True]{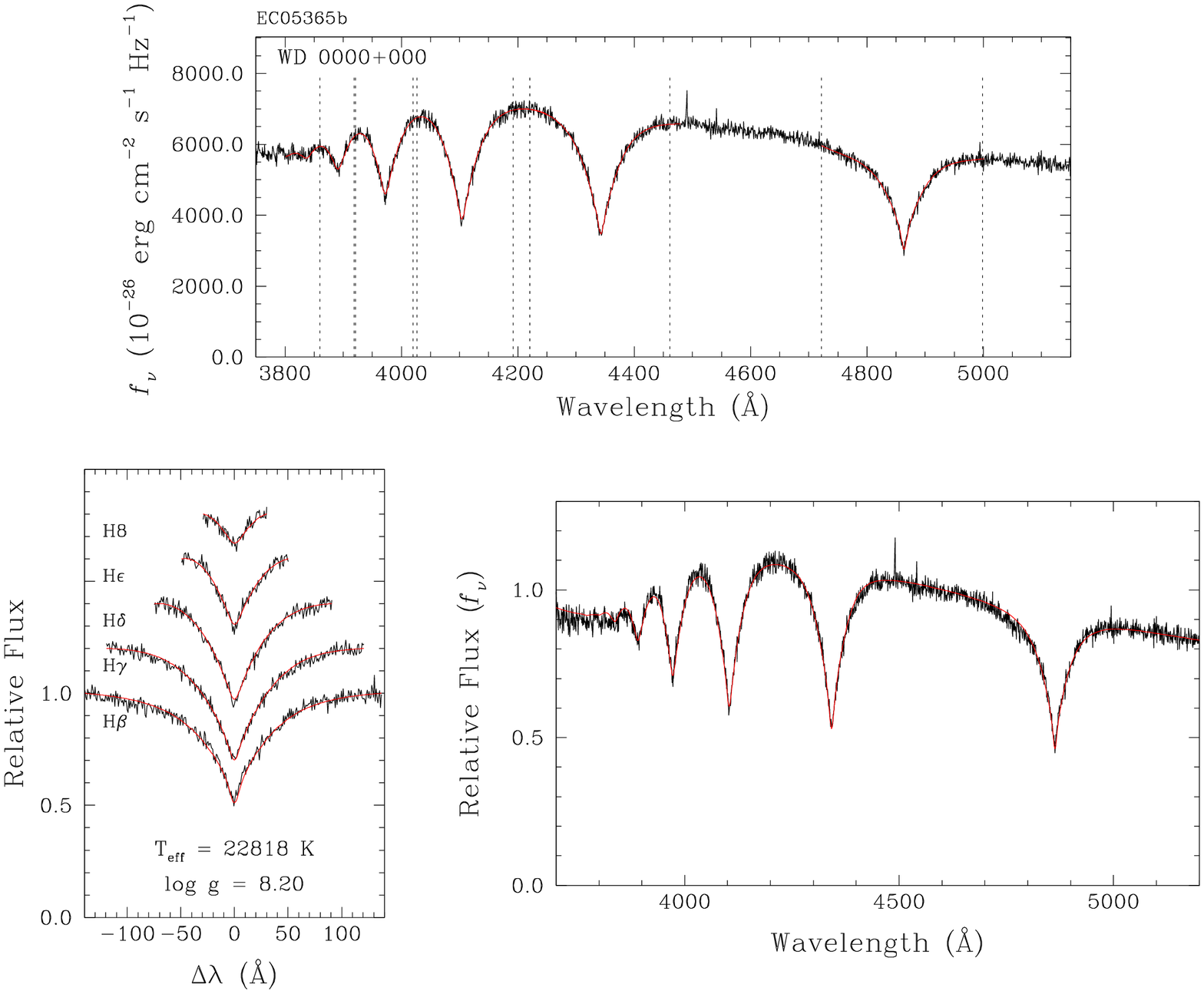}
\caption{Normalized hydrogen balmer lines of spectroscopic data taken with SOAR with best fitted $\log {g}$ and T$_{\rm eff}$ model overplotted in red.}
\end{figure}

Though the atmospheric parameters of white dwarfs can be constrained with photometric data \citep{ber95a}, the surface temperature implied by our photometry is poorly constrained (21\,000-26\,000\,K), which is often the case for stars this hot. We instead rely on the spectral line fitting technique initially developed by \cite{ber92} and refined by \cite{ber95b, lie05}. The method generates a grid of model spectra over a range of surface gravities ($\log {g}$; $g$ measured in cm\,s$^{-2}$) and effective surface temperatures (T$_{\rm eff}$ degrees K). A Levenberg-Marquardt chi-squared minimization is then used to determine which values of $\log{g}$ and T$_{\rm eff}$ best reproduce the Balmer profiles relative to their local continuum. By using only the shape and depth of the Balmer profiles, this method is expected to be robust against overall flux calibration errors. 

We first obtained atmospheric parameters using the low resolution Goodman data and the corresponding spectrophotometric standard data for flux calibration. The best-fit $\log{g}$ and T$_{\rm eff}$ were 8.03 $\pm$ 0.05 and 24\,050 $\pm$ 150\,K with the error bars representing the statistical uncertainties of the fitting process. As expected, the individual Balmer profiles were reproduced well by the model atmosphere but the continuum shape between lines was poorly fitted. Given the cloudy conditions of the night, we suspected the data used for flux calibration to be unreliable. 

To asses the reliability of the standard flux calibration, we generated a comparison set of spectra which were flux calibrated with an independent set of white dwarf model spectra, kindly provided by D. Koester (see \cite{koe10} for details on model input  physics and methods). We chose models spanning the photometrically constrained temperature range and only used continuum points for calibration to avoid biasing the fits to the Balmer profiles. When fitting the model flux calibrated spectra of EC\,05365 we found the best fitting $\log{g}$ and T$_{\rm eff}$ to be 8.20 $\pm$ 0.05 and 22\,800 $\pm$ 150\,K independent of the white dwarf model used for flux calibration. The agreement of the model flux calibrated solutions further supports our suspicion of the standard flux calibrated data. We therefore adopt T$_{\rm eff} = 22\,800^{ +1400}_{ -150}$\,K and $\log{g} = 8.19^{+0.05}_{-0.20}$ for the remainder of the analysis, incorporating the solutions from our standard flux calibrated data into our reported uncertainty. Figure 2 shows the Balmer series fits to the model calibrated Goodman data. Using the evolutionary sequences of \citet{fon01}, the surface gravity and temperature are consistent with a white dwarf mass and cooling age of $0.74^{+.03}_{-.12}$\,M$_{\odot}$ of $63^{+7}_{-36}$\,Myr.

\subsection{Establishing the Infrared Excess}

First, we converted all measured photometry to units of flux density using published zero points \citep{wri10,skr06,hol06}. We scaled the white dwarf model to the observed photometry by calculating a median scale factor for all photometry points without clear excess ($\leq$ 2.15\,$\mu$m). The scaling factor is consistent with a photometric distance of $87^{+13}_{-3}$\,parsecs. The resulting scaled white dwarf model is plotted with the observed photometry in Figure 3. 

\begin{figure}
\begin{centering}
\includegraphics[scale=0.50,clip=True]{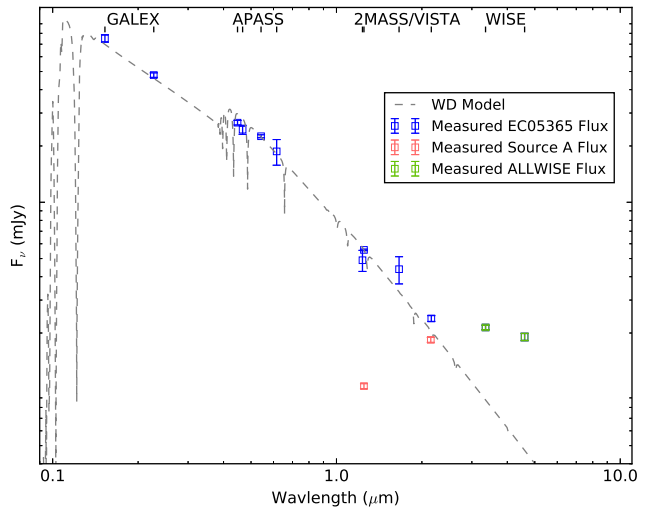}
\end{centering}
\caption{EC$\,$05365 observed photometry in blue overplotted with EC$\,$05365 white dwarf model shown in light grey. The VISTA photometry for Source A and the spatially coincident ALLWISE photometry are plotted separately in red and green.Note that the errors presented are catalog uncertainties and likely underestimate the absolute (including systematic) photometric uncertainty.}
\end{figure}

The departure of the \emph{W1} and \emph{W2} photometry from the white dwarf model of EC$\,$05365 is evidence for an external infrared bright source. But before we can attribute it to a source coincident with EC$\,$05365 we must consider the contribution to the \emph{W1} and \emph{W2} measurement of the nearby extra-galactic Source A. Since there were no photometric flags indicating profile deblending was not performed for this detection, and any contributions from Source A are included in the catalog measurement. 

We are unable to model the expected contribution of the galaxy to \emph{W1} and \emph{W2} bands as it is only reported in the two VISTA bands and we cannot constrain both the spectral type and redshift. Instead we used the galaxy SED atlas of \citep{bro14} to determine whether any nearby galaxy SEDs could match the observed J-Ks colors and reproduce the entire \emph{W1} and \emph{W2} excess. The atlas constitutes a large set of observed SEDs of nearby (z $<$ 0.05) galaxies spanning a wide range of morphological types. Few of the observed catalog SEDs are able to reproduce the observes Source A slope from J to Ks and none of them are likewise consistent with the \emph{W1} and \emph{W2} excess. We are confident this rules out the possibility of a nearby galaxy accounting for the excess, but cannot rule a higher redshift source.

A by-eye extrapolation of the flux from Source A in Figure 3 might suggest a very strong contribution to the \emph{W1} and \emph{W2} measurements, but the spectral energy distribution does not consider all of the information available, namely the relative separations of the ALLWISE photocenter from the measured positions of EC$\,$05365 and Source A in the VISTA VHS catalog. Based on the additional astrometric evidence, we now argue that even if there is contamination from Source A in the \emph{W1} and \emph{W2} measurements, there is still a significant infrared excess associated with EC$\,$05365.

\begin{figure}
\begin{centering}
\includegraphics[scale=0.5,clip=True]{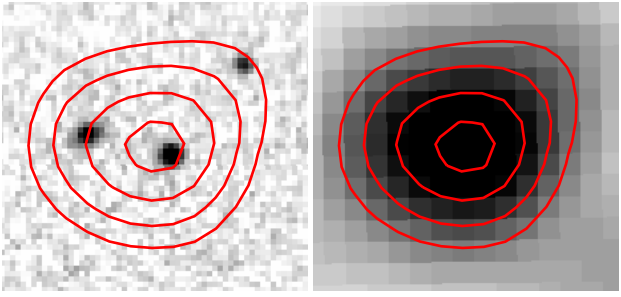}
\end{centering}
\caption{Left: Vista \emph{K$_{s}$} image with \emph{W1} isoflux contours overlaid. Right: \emph{WISE} W1 image with isoflux contours. The images are 18\arcsec x 20\arcsec to match the scale of Figure 1. Both the minor asymmetry of the contours along the line between EC$\,$05365 and Source A and the offset in the ALLWISE photocenter from the position of EC$\,$05635 suggest potential for source confusion.}
\end{figure}

In Figure 4 we show the VISTA \emph{K$_{s}$} and the WISE \emph{W2} band image, overplotted with isoflux contours of the \emph{W2} image. It is immediately apparent that the contours of the \emph{WISE} detection are more consistent with a source detection at the position of the target EC05365, with a minor asymmetry and offset in the direction of Source A suggesting some measureable contamination. As a first-order estimate of the level of contamination, we can use the relative separations of the target and Source A from the \emph{WISE} photocenter as a rough proxy of their relative flux. Assuming the positions of the target and Source A as reported by the VISTA Science Archive and considering the source position of the ALLWISE detection and its associated error ellipse, we find the ratio of the flux of the target to the flux of Source A within the \emph{WISE} bands to be $4.0/1.0 \pm 0.6$. If we subtract this level of contribution of Source A from the ALLWISE flux measurements (approx. 20\%) we still find an excess above the white dwarf model in the \emph{W1} and \emph{W2} bands of 13$\,\sigma$ and 12$\,\sigma$, assuming the reported photometric errors. Given the significance of the excess, we do not expect any systematic error introduced by assuming the reported photometric errors to invalidate the excess.

\floattable
\begin{deluxetable}{lcc|cccc|ccc|ccc|cc}
\tablecaption{EC$\,$05365 Collected Photometry}
\tablehead{\nodata & \multicolumn{2}{c}{GALEX} & \multicolumn{4}{c}{APASS} & \multicolumn{3}{c}{2MASS} & \multicolumn{3}{c}{VISTA} & \multicolumn{2}{c}{WISE\tablenotemark{a}}} 
\startdata
 & \emph{FUV} & \emph{NUV} & \emph{B} & \emph{g} & \emph{V} & \emph{r} & \emph{J} & \emph{H} & \emph{K$_{s}$} &\emph{J} & \emph{H} & \emph{K$_{s}$} & \emph{W1}\tablenotemark{b} &\emph{W2}\tablenotemark{b} \\
 Flux (mJy)  & 7.51 & 4.78 & 2.66 & 2.44 & 2.24  & 1.87 & 0.49 & 0.44 & -- & 0.556 & -- & 0.239 & 0.141 - 0.171 & 0.147 - 0.152 \\
 $\sigma$ (mJy) & 0.34 & 0.13 & 0.07 & 0.13 & 0.04 & 0.29 & 0.06 & 0.07 & -- & 0.005 & -- & 0.008 & 0.004 & 0.008 \\
 \enddata
\tablenotetext{}{Note: The errors presented are catalog uncertainties and likely underestimate the absolute (including systematic) photometric uncertainty}
\tablenotetext{a}{The range of flux is bound by the difference between Case 1 and Case 2 contamination discussed in section 3.2}
\tablenotetext{b}{The original reported flux values for the potentially contaminated ALLWISE catalog source are .211 and .192 (mJy) respectively}
\end{deluxetable}

As a more informative approach, we followed the technique of `forced' photometry as demonstrated by \cite{lan16}. In essence, the technique relies on forward modeling of the WISE catalog images taking the positions of sources, the point-spread function of the band of interest, and a per pixel noise model as knowns and solving for the most likely combination of source fluxes that can reproduce the observed data \citep{lan16}. 

For our images we use the unWISE coadds produced by \cite{lan14}, which are a publicly available set of \emph{WISE} coadds sampled at the nominal resolution of the detector and optimally combined for the purpose of forced photometry. As our point-spread functions for each \emph{WISE} band, we use a combination of gaussian functions as derived in \cite{lan16} with weight and standard deviation given in their Table 1. As an estimate of the local noise we choose three neighboring regions that are source free in the VISTA \emph{J} and \emph{K$_{s}$} images, and in the unWISE coadded images. Since we are only concerned with the relative flux of two sources, we proceed by generating a grid of fluxes for each source. At each point on the grid we convolve the fluxes  at their source position with the point-spread function of the band of interest, sampled at the resolution of the image, and subtract the result from the unWISE image. We then compare the resulting subtracted image with the local noise estimates using the method of least squares to find the most probable combination of fluxes. We perform this exercise separately on the W1 and W2 images. 

As expected, the result of the forced photometry measurement varies between \emph{W1} and \emph{W2}. For the \emph{W1} band, a target to Source A flux ratio of $2.0/1.0$ best reproduced the image. Whereas for the \emph{W2} band, the best fitting target to Source A flux ratio was $3.39/1.0$. Assuming that Source A contributes this fraction of flux to the ALLWISE detection, if we again subtract the contribution from Source A to the \emph{W1} and \emph{W2} measurements we find the remaining flux above the white dwarf model to represent 8$\,\sigma$ and 11$\,\sigma$ excesses respectively.

\subsection{Modeling the Infrared Excess}

The complete photometry set is collected in Table 2. Note that the errors presented are catalog uncertainties and likely underestimate the absolute (including systematic) photometric uncertainty. Given the uncertainty in our \emph{WISE} contamination estimates and the lack of near-infrared data for modeling, we do not expect inflated error bars to have much effect on our conclusions. We proceed by exploring different astrophysical models that can reproduce the infrared excess associated with EC$\,$05365, in each case considering a minimally (using the results from astrometry alone) and maximally (using the results from the forced photometry) contaminated flux measurement for the \emph{W1} and \emph{W2} bands. For the remainder of the section we will refer to these results as Case 1 and Case 2 levels of contamination.

The simplest model we explored was a spherical blackbody source taken to be at the photometric distance of the white dwarf, e.g. an unresolved low mass companion. A minimum chi-square fitting method was used to determine a best fitting temperature and radius for the Case1 and Case 2 contamination levels of 1040\,K, 0.15 R$_{\odot}$  and 800\,K, 0.23 R$_{\odot}$ respectively. The temperatures are consistent with the T dwarf sub-stellar class but the radii are implausibly large. In addition, the \emph{J}-\emph{K$_{s}$} color provided by both 2MASS (\emph{J}--\emph{K$_{s}$} $<$ 0.556) and VISTA (\emph{J}--\emph{K$_{s}$} = --0.28) rule out a companion of this spectral class \citep{cha00}. 

By dividing the bolometric flux of these single temperature blackbody solutions with the bolometric flux of the model white dwarf, we can also get an estimate for the fractional infrared luminosity $\tau = L_{\rm{IR}}/L_{\star}$ of the system. Such a metric has been demonstrated by \cite{far11} and \cite{roc15} to point towards aggregate properties of the discs of dusty white dwarfs. The different levels of contamination result in a range of $\tau = 0.066\--0.080\,\%$, adding EC$\,$05365 to the growing sample of subtle infrared excesses around young white dwarfs.

We also modeled the infrared excess with a flat, optically thick circular disk using the formalism developed by \cite{jur03}. The model invokes a radially dependent temperature distribution and can be used to constrain the inclination, inner, and outer radius of the disk. There is however an unavoidable degeneracy between inclination and the width of the disk, which is amplified in the absence of longer wavelength data. Having only two mid-infrared data points with clear excess, we fixed the inclination and explored only the inner and outer radius of the disk. Considering the degeneracy between width and inclination, we fitted the models under assumptions of both high ($i=80$ deg) and low ($i=45$ deg) inclination, and present both as viable solutions.

\begin{deluxetable}{lcccc}
\tablecaption{Best fitted circular dust disk parameters}
\tablehead{Fixed $i$ & T$_{\rm inner}$ & T$_{\rm outer}$ & R$_{\rm inner}$ & R$_{\rm outer}$ \\  (deg) &  (K) & (K) & (R$_{\star}$) & (R$_{\star}$)}
\tablecolumns{5}
\startdata
\sidehead{Case 1 Contamination:}
45 & 1100 & 1030 & 34 & 37 \\
80 & 1150 & 860 & 32 & 47 \\
\sidehead{Case 2 Contamination:}
45 & 860 & 800 & 47 & 52\\
80 & 920 & 650 & 43 & 68\\
\enddata
\end{deluxetable}

We used a minimum chi-squared fitting procedure to determine the best fit inner radius and disk width. We parameterized the inner radius in terms of its temperature and explored a range of inner temperatures from 500 to 2500\,K and a range of disk widths from 1-20 WD radii. The results from considering both the Case 1 and Case 2 contaminated \emph{W1} and \emph{W2} measurements are summarized in Table 3. We overplot the best fitted combined spectral energy distributions of the white dwarf and the dust disks on the observed photometry for Case 1 and Case 2 contamination in Figure 5. 

The range of parameters than can fit the data is large, but qualitatively the subtle excess forces the circular disk models to two classes of solutions: a wide disk observed at high inclination or a narrow disk observed at low inclination. With no prior expectation for the inclination, there is no preference for either class of solutions. The subtle excess associated with EC$\,$05365 is so far typical of dusty white dwarfs with cooling ages below 200 Myr, as recently explored by \cite{roc15}. The authors demonstrate that the assumption of disks with dust populating the entire available region between the sublimation radius where dust particles are destroyed by the stellar radiation and the tidal disruption radius where the dust is produced is inconsistent with an assumption of randomly inclined disks, given the high inclinations needed to model the subtle infrared excesses commonly observed for younger dusty white dwarfs. The tendency towards narrow disks (which do not span the available space between the sublimation radius and tidal disruption radius) is less apparent for older dusty white dwarfs with cooling ages between 200 and 700 Myr, presenting an interesting delineation between the two samples. 

\begin{figure}
\begin{centering}
\includegraphics[scale=0.48,clip=True]{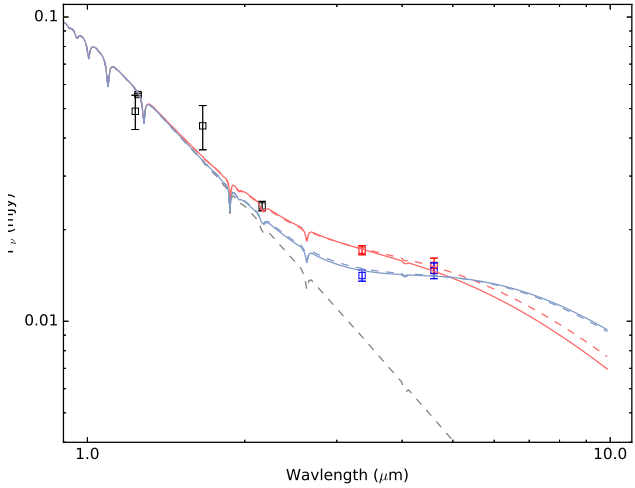}
\end{centering}
\caption{Best-fitted circular disk models overplotted on measured photometry and white dwarf model. The minimally contaminated solutions are shown in red while the maximally contaminated solutions are in blue. The solid/dashed lines represent low/high inclination solutions.}
\end{figure}

As a case study, we explored whether the necessity of narrow disk width for low inclinations could be relieved for EC$\,$05365 by modeling the infrared excess as a set of confocally nested elliptical rings rather than concentric circular rings. This was motivated by the fact that the formation scenario of dusty debris disks necessitates highly eccentric orbits during the tidal disruption phase. The eccentric rings formed by individual disruptions are expected to circularize under the effects of Poynting-Robertson drag on timescales ranging from tens to thousands of years for the relevant particle size scales \citep{ver15}, but the effects of multiple disruption events and collisions on the eccentricity evolution of disks is largely unknown.

\begin{figure}
\begin{centering}
\includegraphics[scale=0.5,clip=True]{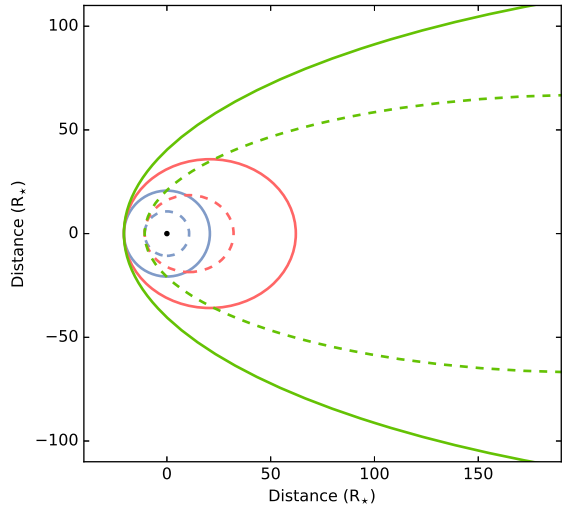}
\end{centering}
\caption{An illustration of geometric differences between disk models with fixed periastron width and increasing eccentricity. The white dwarf is shown as the filled black circle and the blue, red, and light green dashed/solid lines represent inner/outer bounding curves for disks with eccentricities of 0.0, 0.5, and 0.95.}
\end{figure}

\begin{figure*}[t]
\centering
\subfigure{\includegraphics[scale=0.5,clip=True]{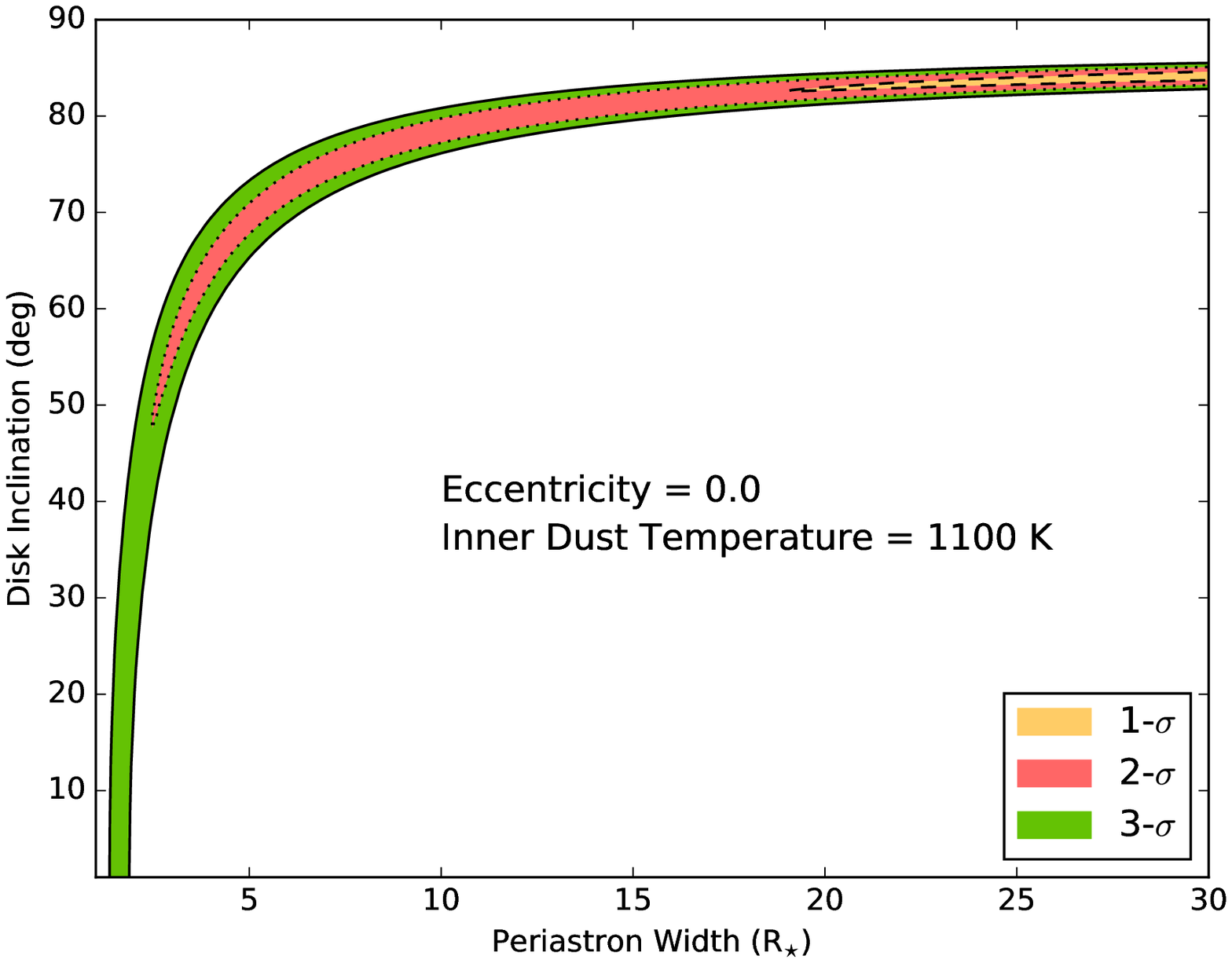}}
\subfigure{\includegraphics[scale=0.5,clip=True]{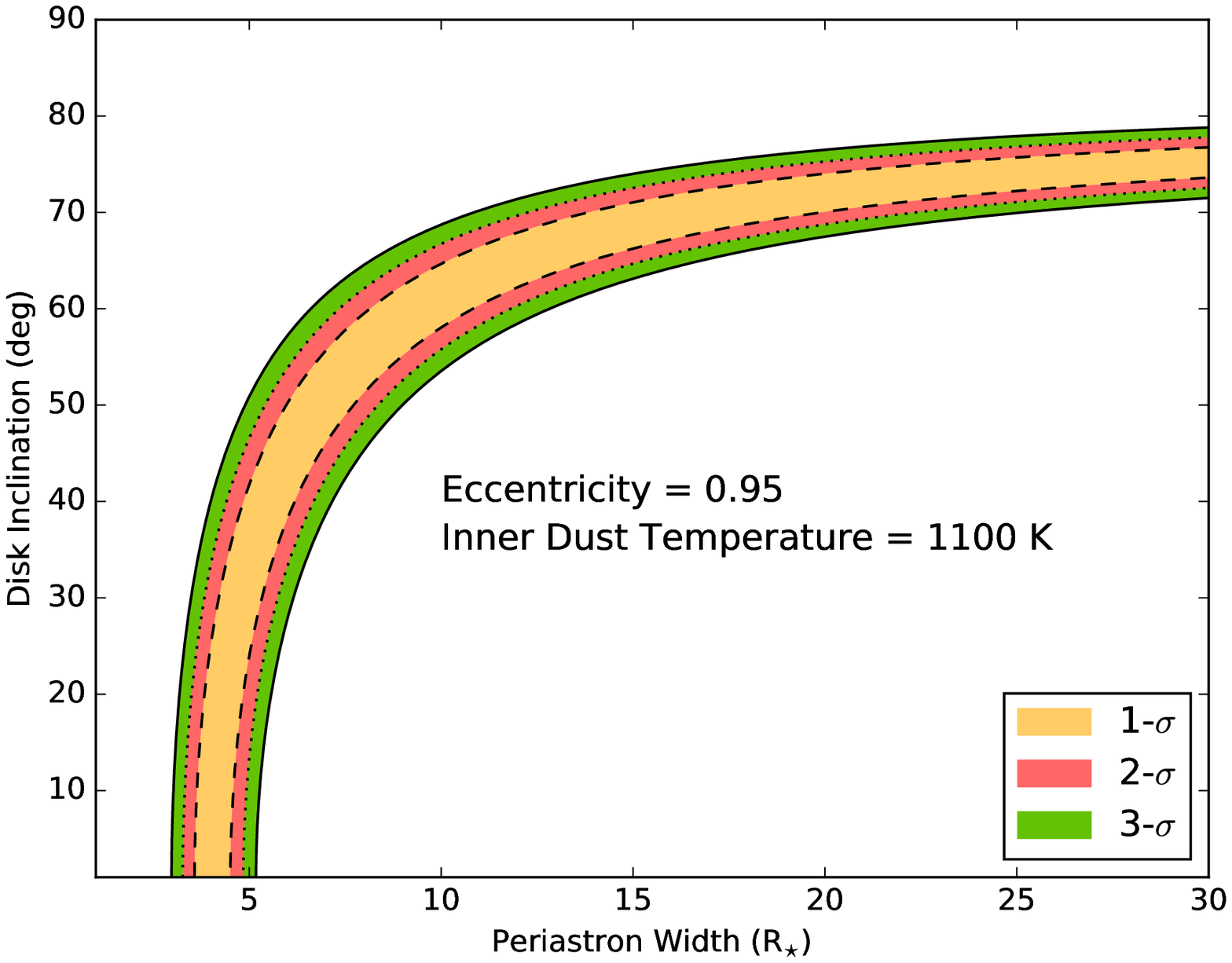}}
\caption{Dashed, dotted, and solid lines show filled contours of 1, 2, and 3-sigma confidence intervals of the periastron width and inclination of a dust disk with inner dust temperature 1100$\,$K around EC$\,$05365. The left-hand panel shows how the near infrared flux of the SED constrains the dust to be in either narrow rings or viewed at high inclinations. The right-hand panel shows how elliptical disks might allow a disk at periastron to be wider over a larger range of inclinations.} 
\end{figure*}

We begin by adopting the same passive, flat, and opaque assumptions about the dust particles as considered by \cite{jur03}. We consider the white dwarf to be at the center of our radial coordinate system, and the inner and outer ellipses that bound the disk to be described by the parameters $a_{\rm in}$, $a_{\rm out}$, and $e$ where $a_{\rm in}$ and $a_{\rm out}$ define the semi-major axes of the inner and outer ellipses, and $e$ defines the eccentricity of the ellipses which we hold fixed for all nested ellipses. The ellipses are confocally nested with the white dwarf at one focus. An illustration of this geometry is shown in Figure 6. In this way, the coordinate $r$ specifies the distance to a dust particle from the white dwarf which uniquely determines the particle temperature under the assumption of an optically thick, flat disk. The inner/outer ellipses which bound the disk are then described by the equation:
\begin{equation}
r_{\rm in/out}(\theta) = \frac{a_{\rm in/out}(1-e^2)}{1-e\cos\theta}
\end{equation} 
and our integral for the observed monochromatic flux as a function of frequency becomes:
\begin{equation}
F_\nu = \frac{\cos i}{D^2}\int_{0}^{2\pi}d\theta \int_{r_{\rm in}(\theta)}^{r_{\rm out}(\theta)} B_\nu\left(T_{\rm ring}(r)\right)rdr
\end{equation}
where $i$ is the inclination of the disk and $D$ is the distance of the system. 

For a qualitative comparison between circular and elliptical models, we applied both models to the Case 2 contamination excess observed for EC$\,$05365 with a fixed inner radius of the disk corresponding to a temperature of 1100K. We varied the difference between the periastron of the inner/outer ellipses and the inclination for both a circular ($e=0.0$) and highly elliptical ($e=0.95$) disk model to produce the constant chi-square boundaries shown in Figure 7. The 1, 2, and 3-$\sigma$ confidence intervals correspond to the chi-squared contours of 2.30, 6.18, and 11.8 above the minimum chi-squared value for each grid, as appropriate for two parameters of interest. The circular and highly eccentric models have comparable minimum chi-squared values which are consistent with a reasonable fit. 

There is a stark difference between the acceptable parameter space for EC$\,$05365 for circular and elliptical disks illustrated by Figure 7: for a given inclination, the elliptical disk is markedly wider at periastron. This can be intuitively understood as close-in material is re-distributed to farther distances along the eccentric orbit, effectively shifting flux from the near-infrared out to longer wavelengths.

This qualitative difference holds true independent of the inner dust temperature assumed. Given the lack of long wavelength data where the models begin to diverge, there is no quantitative (e.g. goodness of fit measurement) distinction between either model, especially considering the addition of the free parameter $e$. The results are however encouraging for elliptical SEDs as an alternative explanation for the lack of younger disks with dust spanning the available orbital range between the sublimation and tidal disruption radius. The delineation between the dust distributions of the younger and older samples could instead be the product of a third parameter, eccentricity. We explore this possibility and the application of elliptical dusty rings to the larger dusty white dwarf sample in a forthcoming paper. 

\subsection{Abundance Measurements and Ratios}

We followed the procedures detailed in \cite{duf10} to estimate the atmospheric abundance of Ca and Mg. In short, assuming the atmospheric parameters above, we generated model white dwarf atmospheres using the codes developed by \cite{duf07} and performed a chi-squared minimization over a grid of atmospheric abundances for each element detected. We find the atmospheric metal pollution to be consistent with logarithmic abundances of [Mg/H] = --5.36\,$\pm$\,0.25 and [Ca/H] = --5.75\,$\pm$\,0.25. Figure 8 shows the best fit profile to the Mg 4481$\,$\AA$\,$ doublet. 

The observed abundances can be transformed into estimates of the accretion rate with knowledge of the rates of diffusion for each element. Using the atmospheric parameters derived in section 3.1 and the updated diffusion calculations described in \cite{koe09}\footnote{See updated tables here: http://www1.astrophysik.uni-kiel.de/~koester/astrophysics/astrophysics.html}, we determined the diffusion timescales for Ca and Mg to be 2.75$\times 10^{-3}$ and 4.47$\times 10^{-3}$ years respectively. The atmospheric calculations of \cite{koe09} also include an estimate of the hydrogen mass necessary to translate the observed photospheric number densities into mass densities for the accretion rates of each species. Assuming the hydrogen mass to an optical depth of  $\tau$\,=\,$2/3$  is $1.73\times10^{-17}$ M$_{\odot}$, we find the system to be accreting Ca at a rate of $3.8\times10^7$ g\,s$^{-1}$ and Mg at rate of $2.6\times10^7$ g\,s$^{-1}$. We did not include the effects of radiative levitation on the observed abundance and inferred accretion rate as they are are less important for lighter elements such as Ca and Mg, and negligible for accretion rates as high as observed \citep{cha10}.

\begin{figure}
\begin{centering}
\includegraphics[scale=0.5,clip=True]{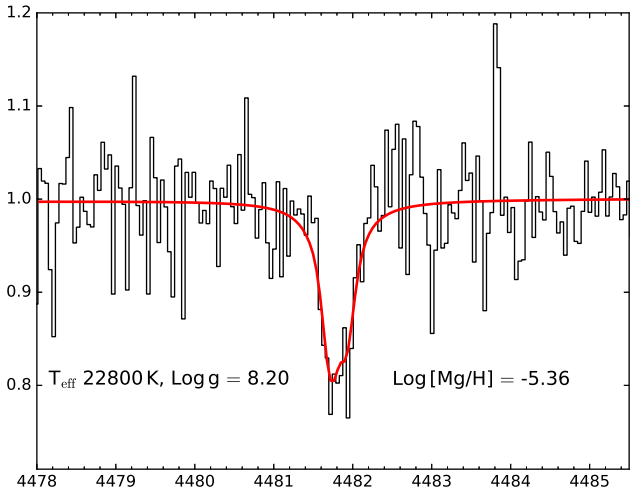}
\end{centering}
\caption{Magellan/Mike data overplotted with the atmospheric abundance model around \ion{Mg}{2} 4481$\,$\AA. The red line is the best fit model with a logarithmic abundance of [Mg/H] = --5.36.}
\end{figure}

The most exciting feature of the dusty metal polluted white dwarfs is the opportunity to translate the observed abundances into the abundances of the accreted parent body, thereby probing rocky exoplanetary abundances not accessible to other direct methods. The observed abundances depend on the rates of accretion, diffusion, and critically the understanding of the current accretion phase. As discussed in \cite{koe09}, the rate of diffusion sets the relevant timescale for the accretion phase and it is reasonable to expect the atmospheric abundances have reached a steady state as long as the accretion has been held constant for \textgreater\,5\,$\tau_{\rm diff}$. In the case of EC\,05365, the diffusion timescales for both Ca and Mg are on the order of days, guaranteeing that accretion has proceeded for several times the diffusion time. Therefore, we can be certain the atmospheric abundance has reached a steady state and, to within the ratios of the diffusion timescales, the observed atmospheric abundance ratios directly reflect that of the accreted parent body. Including the ratio of the diffusion timescales derived above we calculate the parent abundance ratio as $[\rm{Mg}$/$\rm{Ca}] = +0.24 \pm 0.25$.

Because these two elements have strong transitions in the optical, the [Mg/Ca] ratio has been studied for a large sample of polluted white dwarfs (see the recent assessment \cite{jur13}\footnote{\cite{jur13} use the instantaneous accretion approximation (also described as the early phase approximation as it does not account for differences in elemental diffusion timescales) for the sample in their Figure 1. Under this assumption log$[\rm{Mg}$/$\rm{Ca}]= +0.39$ for EC$\,$05365.} in particular their Figure 1). Despite some dispersion, it is clear that the ensemble abundance ratio reflects bulk earth composition, but the abundance ratio for EC$\,$05365 represents a large departure from mean of this sample.

One possibility for this departure could be the accretion of some refractory dominated material, such as the crust of a larger, differentiated body. Such a scenario is discussed by \cite{jur15}, and the spread in abundances ratios of well-studied polluted white dwarfs is further evidence for the post-nebular processing of extrasolar asteroids. The closest analog to EC$\,$05365 is GD$\,$362, which \cite{xu13} demonstrated must have accreted from either multiple distinct sources, or some material that underwent post-nebular processing. A detailed abundance study will be necessary to confirm whether EC$\,$05365 is indeed accreting from such a unique source.

\section{Context and Follow-up}

As the sample of dusty white dwarfs grows, it is important to consider the discovery of each system in the context of the larger dusty white dwarf sample. EC\,05365 is among the hottest and therefore youngest by cooling age of the dusty white dwarfs. The disk parameters are not well constrained but the fractional infrared luminosity $L_{\rm{IR}}/L_{\star}$ = $0.07\%$ is quite low \citep{roc15}, adding to the sample of subtle infrared excesses around young white dwarfs. We have demonstrated for EC$\,$05365 that the application of an elliptical ring model can help relieve the requirement that the subtle excess be modeled by either narrow or highly inclined circular dust rings, and will consider the application of elliptical rings to the larger sample of dusty white dwarfs in a forthcoming paper. 

Though the abundance ratio of $[\rm{Mg}]$/$[\rm{Ca}]$ suggests the accreted material is not strictly bulk earth composition, for comparison with the larger sample of known accreting white dwarfs we scale the accretion rate of Ca assuming bulk earth abundances and determine a total metal accretion rate of $\dot{\rm M}_{\rm Z} = 2.37\times10^9$ g\,s$^{-1}$, consistent with similarly young systems \citep{far12}.

\begin{figure}
\begin{centering}
\includegraphics[scale=0.45,clip=True]{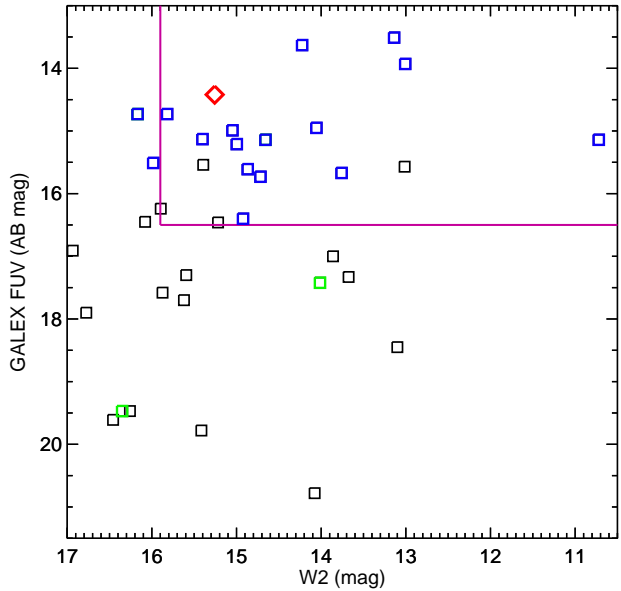}
\end{centering}
\caption{A plot of all currently known dusty WDs with approximate COS and MIRI magnitude limits. Blue squares are WDs already observed with COS while green squares are WDs that have estimated \emph{FUV} magnitudes from their known T$_{\rm eff}$. The red diamond is EC05365. We also over plot rough magnitude limits in purple lines for objects that can have moderate to high resolution UV spectroscopy (S/N $>$ 20 in 2hrs with COS G130M/1291) and moderate resolution mid-infrared spectroscopy with \emph{JWST} (S/N $>$ 10 in 10ks with MIRI/MRS).}
\end{figure}

Detailed abundance studies of highly polluted dusty white dwarfs are the most sensitive probes of rocky exo-
 planetary systems. The combination of optical spectroscopy with large ground-based telescopes and ultra-violet spectroscopy with the Cosmic Origins Spectrograph has proven effective at probing the basic atomic abundances of silicate dust that presumably participated in early terrestrial planet formation around the progenitors of these dusty white dwarfs. More broadly, hot DA white dwarfs with strong mid-IR excesses are particularly valuable in validating the connection between dust abundance, dust accretion, and our understanding of white dwarf photospheric diffusion process. Figure 9 presents a collection of dusty white dwarfs arranged by both \emph{GALEX} \emph{FUV} magnitude and \emph{WISE} \emph{W2} magnitude.  The rough magnitude limits for spectroscopy with COS in the \emph{FUV} or MIRI in the mid-Infrared identify 18 white dwarfs that can have detailed photospheric abundances from the UV, and will also be amenable for detailed mineralogy of the dust residing in their disks via JWST. These two measurements provide independent confirmation of the abundance of the dust in the white dwarf photosphere, and that contained in the disk, such as what has previously been done only for G29-38 \citep{xu14}. We identify this sample, which now includes EC$\,$05365, as critical for future space based follow-up studies.

\acknowledgments

We would like thank the anonymous referee for their detailed reports which greatly improved this manuscript. We would also like to thank Detlev Koester for graciously providing his white dwarf atmospheric models. E. Dennihy and J. C. Clemens acknowledge the support of the National Science Foundation, under award AST-1413001. This work is based on data obtained from (1) the Wide-Field Infrared Survey Explorer, which is a joint project of the University of California, Los Angeles, and the Jet Propulsion Laboratory (JPL), California Institute of Technology (Caltech), funded by the National Aeronautics and Space Administration (NASA); (2) the Two Micron All Sky Survey, a joint project of the University of Massachusetts and the Infrared Processing and Analysis Center (IPAC)/Caltech, funded by NASA and the National Science Foundation (NSF); (3) the 6.5 m Magellan Telescopes located at Las Campanas Observatory, Chile; (4) the Southern Astrophysical Research (SOAR) telescope, which is a joint project of the Minist\'erio da Ci\^encia, Tecnologia, e Inova\c{c}\~{a}o (MCTI) da Rep\'ublica Federativa do Brasil, the U.S. National Optical Astronomy Observatory (NOAO), the University of North Carolina at Chapel Hill (UNC), and Michigan State University (MSU); (5) the VizieR catalog access tool, CDS, Strasbourg, France; (6) the NASA/IPAC Infrared Science Archive, which is operated by JPL, Caltech, under a contract with NASA; (7) the NASA Astrophysics Data System; (8) observations made with the ESO Telescopes at the La Silla or Paranal Observatories under programme ID(s) 179.A-2010 

\facility{WISE, CTIO:2MASS, AAVSO (APASS), Magellan:Clay (MIKE), SOAR (GOODMAN)}

\end{document}